\documentclass[twocolumn,showpacs12,preprintnumbers,amsmath,amssymb,superscriptaddress]{revtex4}

\usepackage{graphicx,epsfig}
\usepackage{bm}

\begin{document}

\title{\bf The phase structure of QGP-Hadron in a statistical model using Cornell, Richardson and Peshier potentials}

\author{ R. Ramanathan, Agam K. Jha$^+$, \\  K. K. Gupta$^*$ and S. S. Singh}

\email{akjha@physics.du.ac.in}

\affiliation{Department of Physics, University of Delhi, Delhi - 110007, India  \\ $^+$Department of  physics, Sri Venkateswara College, University of Delhi, Delhi,India\\  $^*$Department of Physics, Ramjas College, University of Delhi, Delhi - 110007, India}

\begin{abstract}
   We study the phase structure of the QGP-Hadron system under quasi-static equilibrium using the Ramanathan et al. statistical model for the QGP fireball formation in a hadronic medium. While in the earlier published studies we had used the Peshier effective potential which is appropriate for the deconfined QGP phase but could be extrapolated to the transition region from the higher momentum regime, in this paper we study the same system using the Cornell and Richardson potentials which are more relevant for the low momentum confinement regime, but could again be extrapolated to the transition region from below. Surprisingly, the overall picture in both the cases are quite similar with minor divergences,(though,the results with the Richardson potential shows a sizable deviation from the other two potentials), thus indicating the robustness of the model and its selfconsistency. The result of our numerical results pertaining to the variation of the velocity of sound in the QGP-Hadron medium with temperature in the various scenarios considered by us, is that, the phase transition seems to be a gentle roll-over of phases rather than a sharp transition of either the first or second order, a result in conformity with recent lattice calculations, but with much less effort.       \\

     
\end {abstract}

\maketitle
 
\section{Introduction}

\large  The formation of QGP droplet(fireball) is one of the most exciting possibility in ultra relativistic heavy ion collisions(URHIC) [1].The physics of such an event is very complicated and to extract meaningful results from a rigorous use of QCD appropriate to this physical system is almost intractable though heroic efforts at lattice estimation of the problem has been going on for quite some time[2]. One way out is to replicate the approximation schemes which have served as theoretical tools in understanding equally complicated atomic and nuclear systems in atomic and nuclear physics in the context of QGP droplet formation.This approach lays no claim to rigour or ab-initio understanding of the phenomenon but lays the  framework on which more rigorous structures may be built depending on the phenomenological success of the model as and when testable data emerges from ongoing experiments.
 
The nucleation process is driven by statistical fluctuations  which produce the QGP droplets in a hadronic phase,the size of the fluctuations being determined by the critical free energy difference between the two phases.The Kapusta et al model (3) uses the liquid drop model expansion for this,as given by

\begin{eqnarray} 
\Delta F = \frac {4\pi}{3} R^{3} [P_ {had}(T,\mu_{B}) - P_{q,g}(T, \mu_{B})] \nonumber \\
  + 4\pi R^{2} \sigma +\tau_{crit} T~ln \biggl [ 1 + (\frac {4 \pi}{3})R^{3}s_{q,g} \biggl].                                                                     \end{eqnarray}   

The first term represents the volume contribution,the second term is the surface contribution where $\sigma$ is the surface tension,and the last term is the so called shape contribution.The shape contribution is  an entropy term on account of fluctuations in droplet shape which we may ignore in the lowest order approximation.The critical radius $R_{c}$ can be obtained by minimising (1) withrespect to the droplet radius R,which in the Linde approximation[6] is,

 \begin{equation}
 R_{c} = \frac {2\sigma}{\Delta p}~ or~                                             \sigma = \frac {3\Delta F_{c}}{4\pi R_{c}^{2}}
\end{equation}

\section{The statistical model}
In the approximation scheme of the Ramanathan et.al [4 ] the relativistic density of states for the quarks and gluons is constructed adapting the procedures of the Thomas-Fermi construction of the electronic density of states for complex atoms and the Bethe density of states[6] for nucleons in complex nuclei as templates .The expression for the density of states for the quarks and gluons(q,g) in this model is:

\begin{equation}\label{3.13}
\rho_{q, g} (k) =(\nu / \pi^2) \lbrace (-V_{conf}(k))^{2}(-\frac{dV_{conf}(k)}{dk}) \rbrace_{q, g}, 
\end{equation}

where $k$ is the relativistic four-momentum of the quarks and the gluons, $\nu$ is the volume of the fire ball taken to be a constant in the first approximation and V is a suitable confining potential relevant to the current quarks and gluons in the QGP [4] given by:

 \begin{equation}\label{3.18}
V_{\mbox{eff}}(k)k^{3} = (1/2k)\gamma_{g,q} ~ g^{2} (k) T^{2} - m_{0}^{2} / 2k~.
\end{equation}
                    
where $m$ is the mass of the quark which we take as zero for the up and down quarks and $150~MeV$ for the strange quarks.The g(k) is the QCD running coupling constant given by

\begin{equation}\label{3.17}
g^2(k) =(4/3)(12\pi/27)\lbrace 1/ \ln(1+k^{2}/\Lambda^{2}) \rbrace ~,
\end{equation}

where $\Lambda$ is the QCD scale taken to be $150~ MeV$.In fact, eq.(3) may be considered as the starting ansatz of the model without further elaboration as far as our calculational program is concerned.  

The model has a natural low energy cut off at:

\begin{equation}\label{3.19}
k_{min}=V(k_{min})~or~ k_{min}=(\gamma_{g,q}N^{\frac{1}{3}} T^{2} \Lambda^2 / 2)^{1/4},
\end{equation}

with $N=[(4/3 )(12 \pi / 27)]^{3}$.

The free energy of the respective cases i(quarks,gluons,interface etc.) for Fermions and Bosons(upper sign or lower sign) can be computed using the following expression:

\begin{equation}\label{3.20}
F_i = \mp T g_i \int dk \rho_i (k) \ln (1 \pm e^{-(\sqrt{m_{i}^2 + k^2}) /T})~,
\end{equation}

With the surface free-energy given by a modified Weyl[7] expression:
\begin{equation}\label{3.21}
F_{interface} = \gamma T \int dk \rho_{weyl} (k) \delta (k-T)~,
\end{equation}

where the hydrodynamical flow parameter for the surface is:

\begin{equation}\label{3.22}
\gamma = \sqrt{2}\times \sqrt{(1/\gamma_{g})^{2} + (1 / \gamma_{q})^{2}},
\end{equation}

For the pion which for simplicity represents the hadronic medium in which the fire ball resides ,the free energy is:

\begin{equation}\label{3.25}
F_{\pi} = (3T/2\pi^2 )\nu \int_0^{\infty} k^2 dk \ln (1 - e^{-\sqrt{m_{\pi}^2 + k^2} / T})~.
\end{equation}

With these ingredients we can compute the free-energy change with respect to both the droplet radius and temperature to get a physical picture of the fireball formation,the nucleation rate governing the droplet formation, the nature of the phase transition etc. This can be done over a whole range of flow-parameter values [4], We exhibit only the two most promising scenarios in fig.$1$ and fig.$2$. Further  investigating the properties of the corresponding free energies as in [4], it was found that the QGP-hadron phase transition is a weakly first order one.The figs. indicate that the set of flow parameter values used in fig. 1 lead to more stable droplets for a whole range of temperatures with experimentally significant fireballs of radius of the order of $5~fermi$. The increase in the barrier height with temperature is indicative of the fact that the droplets will be more stable at higher temperatures. However, as the rate of droplet formation is controlled by the ratio of the barrier height to temperature, the probability of having more stable droplets is also smaller at higher temperatures. The $S~Vs.~T$ graphs [4] by their smoothness indicate that the transition for a finite chemical potential is neither a first order nor a second order one, but a roll over of phase from the hadronic to the QGP.   

We could have used any potential form in our model instead of the modified Peshier potential (4) used above, the main reason for our use of this form was that the dynamical nature of the quarks in the QGP is well brought out by this. However, there is also a widespred use of the Cornell and Richardson potentials in the recent literature [9, 10] in the context of QGP, though there is also misgiving in some quarters about the validity of using these potentials defined mainly for static constituent quarks in the QGP system. For the sake of testing the robustness of our statistical model we have done the calculations using the Cornell potential [9] and Richardson Potential [9a] in our formalism.  

For Gluon Sector, the cornell potential in momentum space is 

\begin{equation}
V(k)=\frac{12\pi\alpha_{s}}{k^{2}}+\frac{8\pi C \sigma}{k^{4}}
\end{equation}

where $\alpha_{s}$=Strong coupling constant 
                  =$0.09$,
              
              C = Casimir Scaling 
                 = $9/4$, 
      $\sigma$ = String tension
              = 8$T^{2}_{c}$, $T_{c}=170~MeV$
For Quark Sector, the Cornell potential for three flavors in momentum space is

\begin{equation}
V(k)=\frac{16\pi\alpha_{s}}{3k^{2}}+\frac{8\pi \sigma}{k^{4}}
\end{equation}

where $\alpha_{s}$= $0.009$,
      $\sigma$ = $15T^{2}_{c}$, $T_{c}=170~MeV$

For the Richardson potential in the momentum frame we use the form [9a]

\begin{equation}
V(k)=(12\pi /27)(4/3)(1/k^{2})(1/ \ln (1+k^{2}/\Lambda^{2}))
\end{equation}

 where $\Lambda=150~MeV$ and $k$ is four momentum.

We can thus redo our calculations using the above potential forms in our formulation, using the parameter values widely used in the literature.

\section{Results and Conclusion}

 The numerical results of our calculations is summarized by the graphs figs.$1$ to $6$, where the temperature variation of the velocity of sound in the QGP-Hadron system is displayed for the typical parameter values of the model for the three different potentials namely the Peshier,the Cornell and the Richardson potentials. 

 \begin{figure}
\epsfig{figure=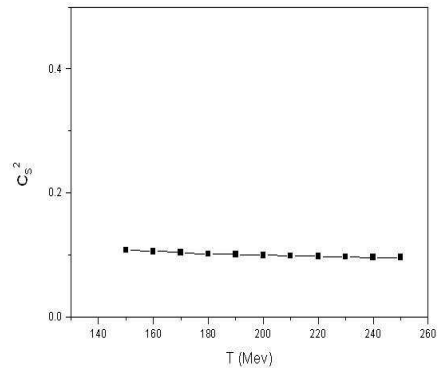,height=2.5in,width=2.5in}
\label{CS2Tmu300.eps}
\caption{\large  Variation of $C_{s}^{2}$ with temperature T at $\gamma_{g} = 6\gamma_{q}$ , $ \gamma_{q} = 1/6 $ and $\mu=300$ with Peshier Potential .}
\end{figure}

 \begin{figure}
\epsfig{figure=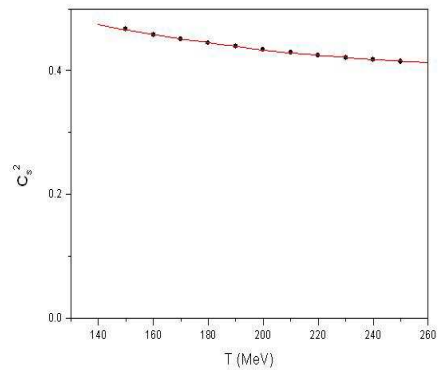,height=2.5in,width=2.5in}
\label{cs2tc.eps}
\caption{\large  Variation of $C_{s}^{2}$ with temperature T with Cornell Potential.}
\end{figure}

\begin{figure}
\epsfig{figure=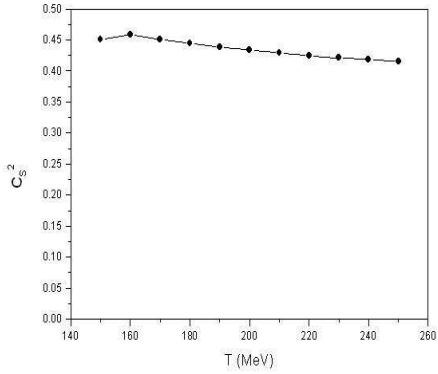,height=2.5in,width=2.5in}
\label{q6.eps}
\caption{\large  Variation of $C_{s}^{2}$ with temperature T at $\gamma_{g} = 6\gamma_{q}$ , $ \gamma_{q} = 1/6 $ and $\mu=0,300$ with Cornell Potential for only three flavored quark sector.}
\end{figure}

\begin{figure}
\epsfig{figure=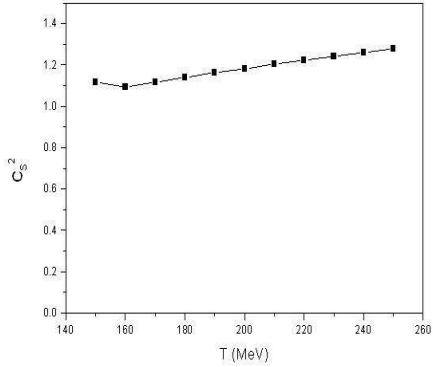,height=2.5in,width=2.5in}
\label{g6.eps}
\caption{\large  Variation of $C_{s}^{2}$ with temperature T at $\gamma_{g} = 6\gamma_{q}$ , $ \gamma_{q} = 1/6 $ and $\mu=0,300$ with Cornell Potential for pure gauge field.}
\end{figure}

 \begin{figure}
\epsfig{figure=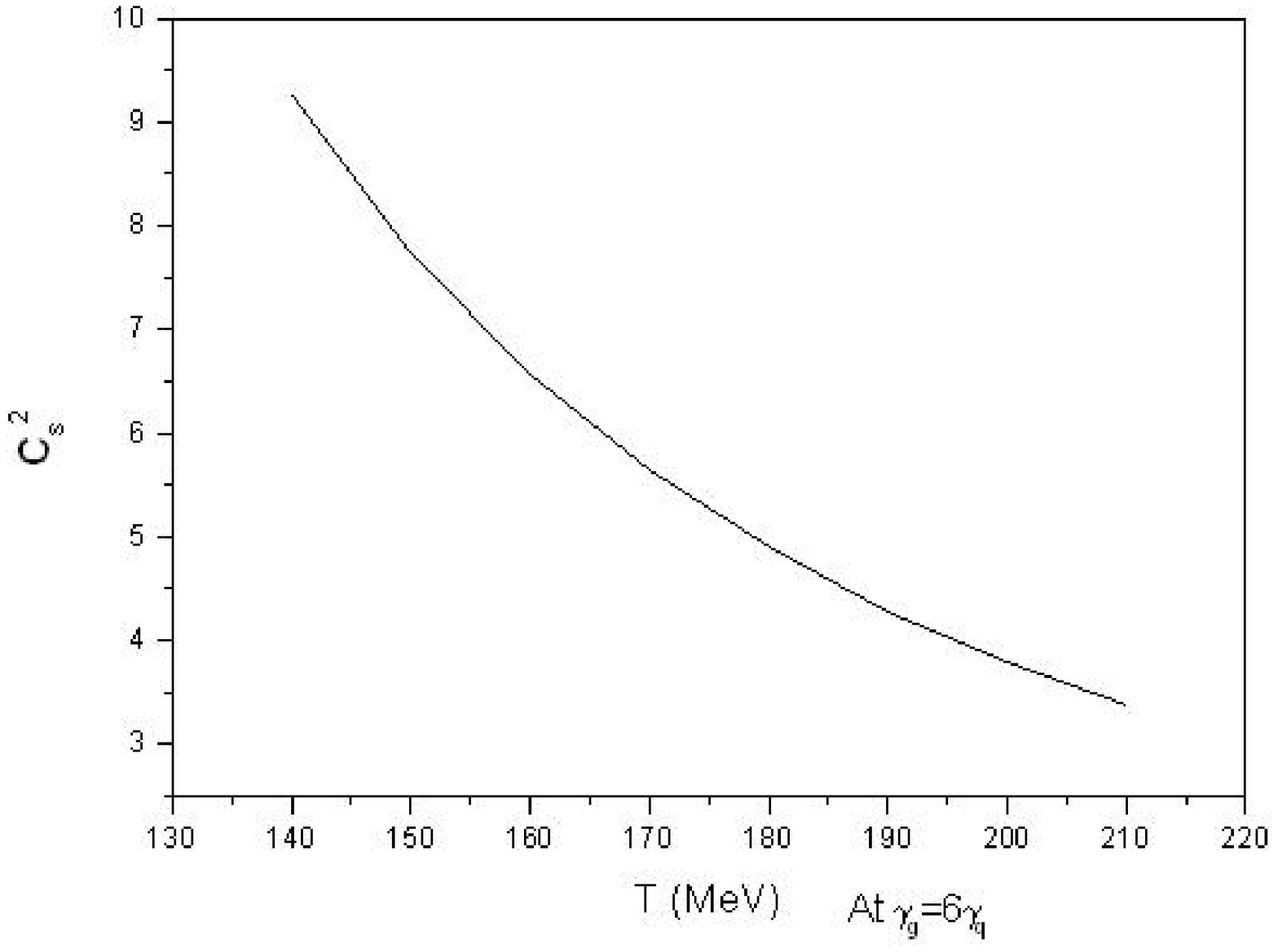,height=2.5in,width=2.5in}
\label{fig.eps}
\caption{\large  Variation of $C_{s}^{2}$ with temperature T with Richardson potetial at $\gamma_{g}=6\gamma_{q}$.}
\end{figure}

\begin{figure}
\epsfig{figure=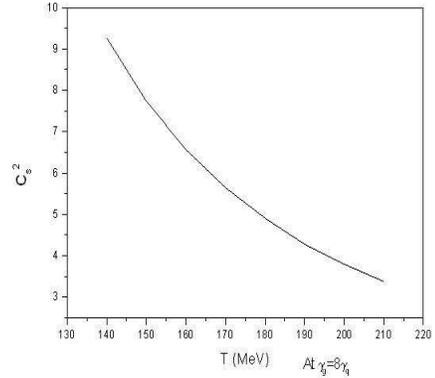,height=2.5in,width=2.5in}
\label{fig1.eps}
\caption{\large  Variation of $C_{s}^{2}$ with temperature T with Richardson Potential at $\gamma_{g}=8\gamma_{q}$.}
\end{figure}

When we compare the numerical results arising from the two different potentials we have considered, the nature of phase transition remains the same in both the potentials' scenarios. The velocity of sound, whose square is the ratio of the entropy and specific heat at constant volume, for the two potentials are not too different as indicated by the graphs (1, 2). While the peshier potential gives almost a contant velocity over a whole range of temperature variation, so does the Cornell potential, except for a distinct kink at around $160~MeV$, the expected critical temperature for the transition.However, in the case of the Richardson Potential, the fall in $C_{s}^{2}$ Vs. $T$ is steeper.So, this difference in their prediction could be used to discard one of the potential forms which disagrees with the actual measurement of sound velocity in a QGP fireball formation event in the foreseeble future.The shift in the absolute values of the velocity in the two cases should not hold much physical significance, as the shift seems to be the result of some overall factor in the potentials used. As we have already noted that we could have used any effective potential available in the literature of quark-gluon interaction, our present excercise only strengthens our confidence in the robustness of the statistical model used as two totally unrelated effective potentials, one, the peshier, designed for the deconfined QGP phase and the Cornell for the confined lower momentum regime lead us ti similar scenarios in the QGP-Hadron transition region.Evidently the ideal gas limit of $C_{S}^{2}=\frac{1}{3}$ is obeyed only by the graph in fig.1, indicating that the Cornell and Richardson potentials lead to violation of this limit, adding to our already stated reasons for doubting the validity of using these potential forms in the analysis of the QGP-Hadron system. The final say in this matter, however, rests with experimental data on the behaviour of $C_{S}^{2}$ in QGP. The overall picture presented by the various graphs, presented here, is the uniform variation of the velocity of sound over a range of temperatures including the expected QGP-Hadron transition temperature of around $170~MeV$, indicating a smooth roll-over of phases rather than a sharp transition. 

{\bf Note:}

To give a brief outline of the origin of eq.(3), which we have already given in detail [4], we recall that the general Thomas-Fermi requirement on the number of particle states in configuration space is given by 

\begin{equation*}
\int_0^{r}\rho(r)d^{3}r=F(V(r))
\tag{3a}\end{equation*}

where $\rho(r)$ is the coordinate-space density of states and $F(V(r))$is a functional of the coordinate-space potential $V(r)$, both sides of (3a) being dimensionless. The momentum space counterpart of (3a) is 

 \begin{equation*}
\int_0^{k}\rho(k)dk=\phi(V(k))
\tag{3b}\end{equation*}

where in this definition of the density of states $\rho(k)$, $k$ is the energy variable and $\phi(V(k))$ is the corresponding transform of $F(V(r))$. In this generalised form (3a), the potential may be either central as in atomic case, or a few body potential as in the nuclear and quark physics. In general to leading order the following forms are appropriate for $\phi(V(k))$, namely 

\begin{equation*}
[\phi(V(k))]_{non-relat.}\approx(Const.)_{non-relat.}[-V(k)]^{\frac{3}{2}}
\tag{3c}\end{equation*}

 \begin{equation*}
[\phi(V(k))]_{relat.}\approx(Const.)_{relat.}[-V(k)]^{3}
\tag{3d}\end{equation*}

We can recover eq. (3) when we differentiate (3b) after substituting eq.(3d) in it with an appropriate choice of the constant dictated by the geometry of the system.

{\bf References :}
\begin{enumerate}
\item{F. Karsch, E. Laermann, A. Peikert, Ch. Schmidt and S. Stickan, Nucl. Phys. B (proc. Suppl.) 94, 411 (2001).}
 \item{T. Renk, R. Schneider, and W. Weise, Phys. Rev. C 66, 014902 (2002).}
\item{L. P. Csernai, J. I. Kapusta, R. Venugopalan, and E. Osnes, Phys. Rev. D  67, 045003 (2003); J. I. Kapusta, R. Venugopalan, and  A. P. Vischer,        Phys. Rev. C 51, 901 (1995); L. P. Csernai and J. I. Kapusta, Phys. Rev. D 46, 1379 (1992). P. Shukla and A. K. Mohanty,  Phys. Rev. C 64, 054910 (2001).}

\item{R. Ramanathan, Y. K. Mathur,  K. K. Gupta, and Agam K. Jha Phys.Rev.C70,027903 (2004); ; R. Ramanathan, ,  K. K. Gupta, Agam K. Jha, and S.S.Singh, Pramana 68,757 (2007).}
\item{B.D.Malhotra and R.Ramanathan,Phys.Lett.A108:153,(1985).}
\item{E. Fermi, Z. Phys. 48, 73 (1928); L. H. Thomas, Proc. Cambridge Philos. Soc. 23, 542 (1927); H. A. Bethe, Rev. Mod. Phys. 9, 69 (1937).}
 \item{H. Weyl, Nachr. Akad. Wiss Gottingen 110 (1911).}
 \item{G.Neergaard and J. Madsen, Phys. Rev. D 62, 034005 (2000).}
\item{K. M. Udayanandan, P. Sethumadhavan, and V. M. Bannur, Phys. Rev. C 76, 044908 (2007);(9a)J. L. Richardson, Phys. Lett. B 82, 272{1979}}
\item{S. Terranova and A. Bonasera, Phys. Rev. C 70, 024906(2004)}
	
\end{enumerate}

\end{document}